\documentclass{DISproc}

\usepackage{amsmath}
\usepackage{amssymb}
\usepackage{amsfonts}

\newcommand{\beq}{\begin{equation}}
\newcommand{\eeq}{\end{equation}}
\newcommand{\bea}{\begin{eqnarray}}
\newcommand{\eea}{\end{eqnarray}}
\newcommand{\bdm}{\begin{displaymath}}
\newcommand{\edm}{\end{displaymath}}
\def\as{\alpha_s}

\def\m{{\cal M}}

\def\ord{{O}}

\def\d{\partial}

\def \d{{\rm d} }
\def \f{f^{\rm gap}}

\begin{document}
\title{Probing colour flow with jet vetoes}

\author{{\slshape Simone Marzani}\\[1ex]
Institute for Particle Physics Phenomenology, Durham University, \\Durham DH1 3LE, United Kingdom}


\maketitle

\begin{abstract}
  We discuss jet vetoes as a means of probing colour flow in hard-scattering processes in hadronic collisions. As an example, we describe a calculation of the dijet cross-section with a jet veto, which resums the leading logarithms of the veto scale and is matched to a fixed-order computation. We compare this prediction to the measurement performed by the ATLAS collaboration.  Finally, we outline future developments in this research area.
\end{abstract}
\section{Introduction}
The phenomenology of strong interactions and, in particular, jet physics are playing a central role in the physics program during these first years of the LHC~\cite{atlasdijet,atlasazim,atlasveto, atlassub,cmsazim,cmsdijet,cmsincl,cms3to2}. In the basic picture of perturbative QCD, hard partons are produced in high-energy collisions and hadronise into colour-neutral particles, which are typically highly collimated into jets. In reality, QCD provides us with a much richer picture, which potentially leads to interesting phenomenology.  Studies of inter-jet radiation and colour connections between jets can unveil non-trivial information about the underlying hard-scattering process. Beyond their own interest as pure QCD effects, these correlations play an important role in many analyses and can be used, for instance, to reduce overwhelming backgrounds.

Jet vetoes appear often in particle physics analyses as tool to keep jet multiplicity fixed. They can also be used to select or suppress certain contributions in specific processes, such as Higgs production in association with two jets, where one can enhance the vector-boson-fusion (VBF) component with respect to the gluon-gluon fusion (GF) one by applying a central jet veto~\cite{DKT, bjorken, minijetveto}. 

Jet vetoes can also be used a means to probe the colour structure of hard processes. For instance, in Refs.~\cite{sung,ttbpheno}, the problem of identifying the colour of a heavy resonance decaying into a $t \bar{t}$ pair was considered. Clearly, the radiation emitted by this new particle depends on its colour charge. However,  this poses difficulties, since the amount of relatively soft radiation is heavily influenced by the underlying event. One can instead study the response of such radiation to the presence of a jet veto and, if the veto scale $Q_0$ is kept large enough, one can minimise contaminations from non-perturbative physics.

Another interesting analysis~\cite{CoxForPilk} suggests studying the cross-section as function of the veto scale for associated Higgs production with two jets  and a jet veto, in order to simultaneously determine the effective coupling of Higgs to the gluons and to the vector bosons. This analysis shows that the biggest theoretical uncertainty comes from the $Q_0$ dependence of the GF channel.

If  jet vetoes are to be used to extract information on the colour flow of hard processes, possibly involving new physics, the $Q_0$ dependence needs to be under good theoretical control.
Unfortunately, performing accurate predictions in perturbative QCD in the presence of jet-vetoes is far from trivial: large logarithms of the ratio of veto scale over the hard scale of the process $Q_0/Q$ contaminate the perturbative expansion and they need to be resummed to all orders (for recent theoretical work see, for instance, Refs.~\cite{BSZveto, BNveto}). Moreover, if the veto is applied only in given regions of phase-space, the observable becomes non-global and the resummation more complex~\cite{nonglobal}.  Monte Carlo parton showers are often used in these studies, but they neglect sub-leading $N_c$ terms, so a better theoretical treatment is needed.

We start  by considering the simplest process, i.e. dijets events, with a veto on the emission of additional radiation in the inter-jet region. This measurement has already been performed~\cite{atlasveto}, so we have been able to use data to validate our theoretical predictions~\cite{FKS,DFMS}.

\section{The dijet cross-section with a jet veto}
We are interested in dijet production in proton-proton collisions at $\sqrt{S}= 7$~TeV, where we veto the emission of a third jet  with transverse momentum bigger than $Q_0 $ in the rapidity region between the two jets. The veto scale is chosen to be $Q_0=20$~GeV.
We define the gap fraction as the ratio of the cross-section for this process over the inclusive rate:
\beq \label{gapfracdef1}
f^{\rm gap} = \frac{\d^2 \sigma^{\rm gap}}{\d Q \d \Delta y}\Big/\frac{\d^2 \sigma}{\d Q \d \Delta y};
\eeq
clearly, in the Born approximation, every event is a gap event and so $\f = 1$. Beyond the Born approximation, the leading jets are no longer balanced in transverse momentum. We choose $Q$ as the mean of the transverse momenta of the leading jets. This choice, in contrast for instance to the transverse momentum of the leading jet, is more stable under the inclusion of radiative corrections. The rapidity separation $\Delta y$ is measured from the centres of the leading jets.

The technique for resumming logarithms of the ratio $Q/Q_0$ for the gaps-between-jets cross section has been known for quite a long time, see for instance~\cite{KOS,SLL1,SLLind}. The observable we are studying is non-global and for this reason ``in-gap'' virtual corrections are not enough to capture the single logarithmic accuracy. Radiation outside the gap is forbidden to re-emit back into the gap, inducing non-global logarithms~\cite{nonglobal}. Currently, these contributions can be resummed only in the large $N_c$ approximation. 
Here, instead, we adopt a different approach: we keep the full colour structure but we expand in the number of gluons, real or virtual, outside the gap; it was argued in~\cite{FKS} that this way of proceeding is a reasonable one. We resum the contributions arising from allowing zero and one gluons outside the gap:
\beq
\frac{\d^2 \sigma^{\rm gap }_{\rm res }}{\d Q \d \Delta y } =\frac{\d^2 \sigma^{(0)}}{\d Q \d \Delta y } +\frac{\d^2 \sigma^{(1)}}{\d Q \d \Delta y }+\dots
\eeq
The first contribution to this expansion is obtained by  considering the original four-parton matrix element, dressed by in-gap virtual gluons, with transverse momenta above $Q_0$. No out-of-gap gluons are included. The resummed partonic cross section has the form
\beq \label{resummedpartonicxsec2}
|\m^{(0)}|^2   =  {\rm tr} \left( He^{- \xi(Q_0,Q) {\Gamma}^{\dagger}}e^{- \xi(Q_0,Q) {\Gamma}}\right), \quad {\rm with }\quad \xi(k_1,k_2)=\frac{2}{\pi} \int_{k_1}^{k_2} \frac{d k_t}{k_t} \as(k_t),
\eeq
where $H$ is a matrix describing the hard-scattering and $\Gamma$ is the soft anomalous dimension matrix, which describes the evolution of a four-parton system.
We also aim to resum the non-global logarithms that arise as a result of allowing one soft gluon outside the rapidity gap. The general framework in which this calculation is performed is described in~\cite{SLL1,SLLind}. We must now consider both real ($\Omega_R$) and virtual ($\Omega_V$) corrections to the four-parton scattering, each dressed with any number of soft gluons:
\beq \label{master1}
|\m^{(1)}|^2 = - \frac{2 }{\pi} \int_{Q_0}^{Q} \frac{d k_t}{k_t}\as(k_t) \int_{\rm out} \left( \Omega_R + \Omega_V\right)\,.
\eeq
In the case that the out-of-gap gluon is virtual, the subsequent evolution is still given by ${\Gamma}$; in the case of real emission, we have to consider the colour evolution of a five parton system.
It has been shown~\cite{SLL1,SLLind} that QCD coherence is violated in this process at sufficiently high perturbative orders, because of Coulomb gluon exchange. As a consequence super-leading logarithms originate when the out-of-gap gluon becomes collinear with one of the initial-state partons. This affects the gap fraction at $\ord(\as^4)$ and beyond. The numerical impact of these contributions has been studied in~\cite{FKS} and was found to be generally modest. 

One of the findings of Ref.~\cite{FKS} was that the gap fraction is under-estimated when it is computed purely with soft gluon techniques. 
In the eikonal approximation energy-momentum is not conserved and there is no recoil of the hard lines against the emissions, which are considered soft. Emissions of ``soft'' gluons  with $k_t \gtrsim Q_0$ do not cost any energy and if they end up in the rapidity region between the leading jets, they spoil the gap.  
Thus, a pure eikonal calculation tends to produce too small a gap fraction. The result can be improved by matching to a fixed order calculation. 
We have performed the leading-order (LO) matching to the full $2\to3$ tree-level matrix elements in Ref.~\cite{DFMS}, where we have also approximately taken into account energy-momentum conservation in the resummation by shifting the argument $x$ of the parton distribution functions.

\begin{figure} 
\begin{center}
\includegraphics[width=0.49\textwidth, clip]{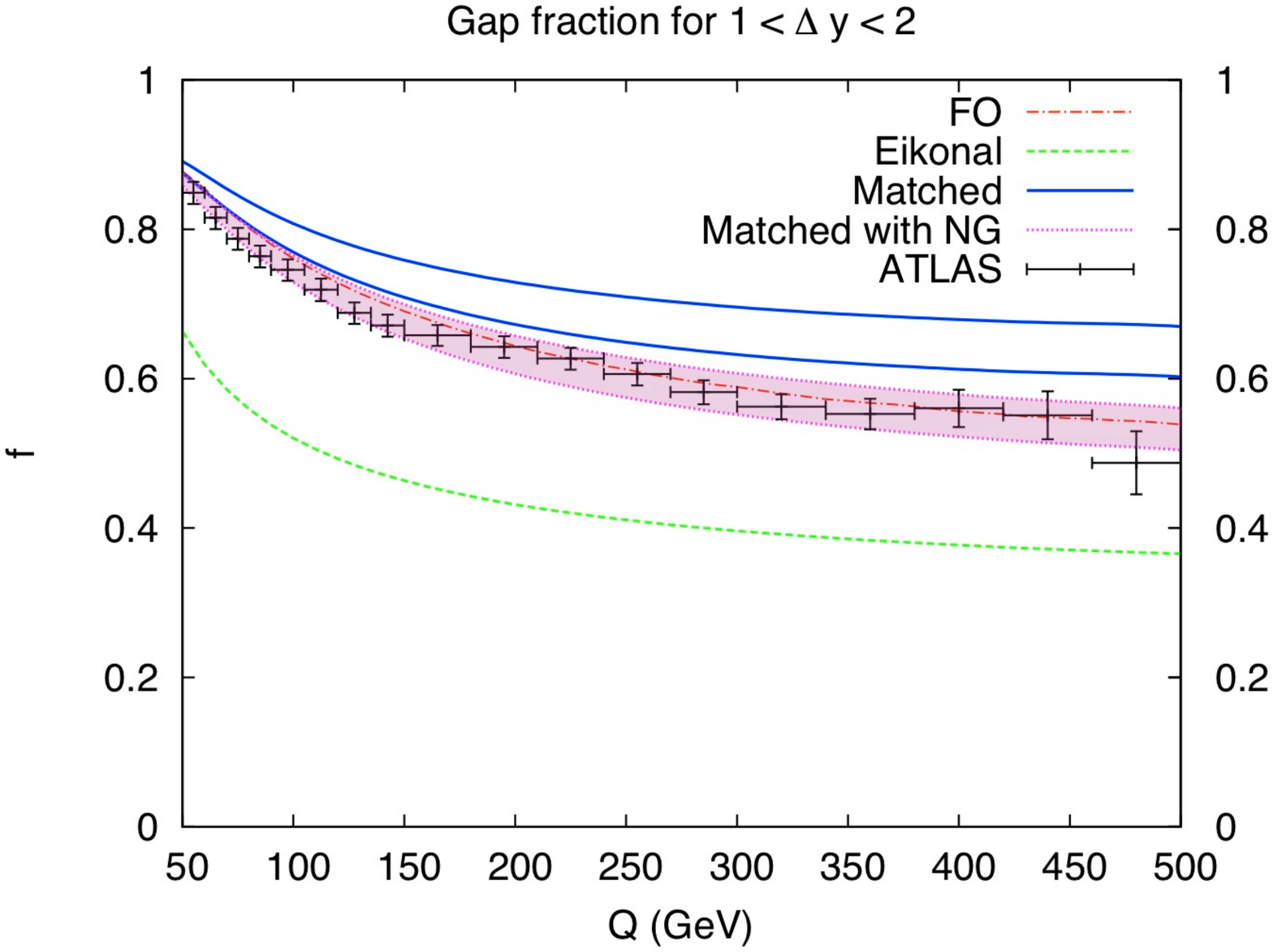}
\includegraphics[width=0.49\textwidth, clip]{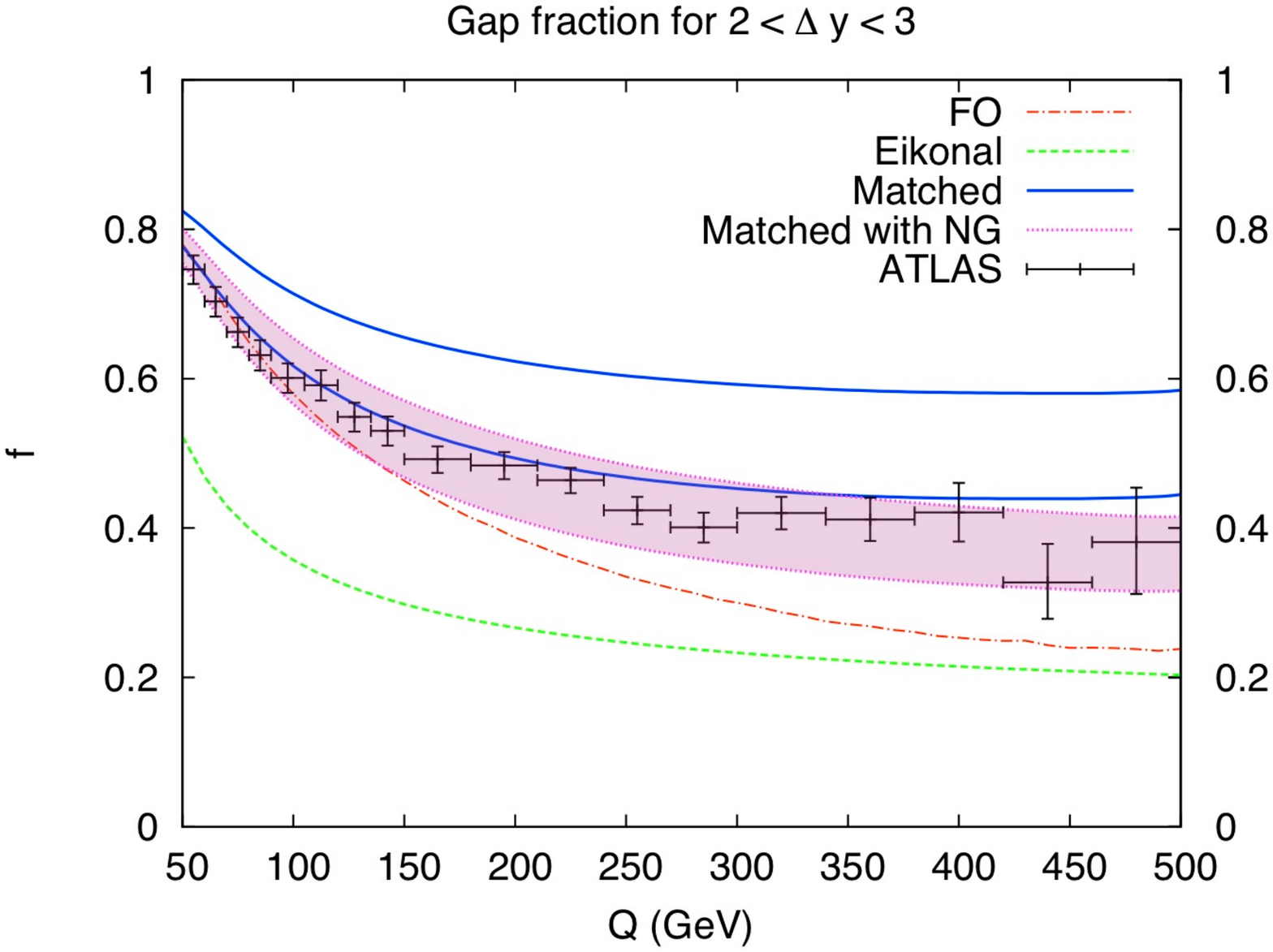}
\caption{The matched gap fraction as a function of the transverse momentum $Q$ in different rapidity bins. }\label{fig:match1}
\end{center}
\end{figure}
\begin{figure} 
\begin{center}
\includegraphics[width=0.49\textwidth, clip]{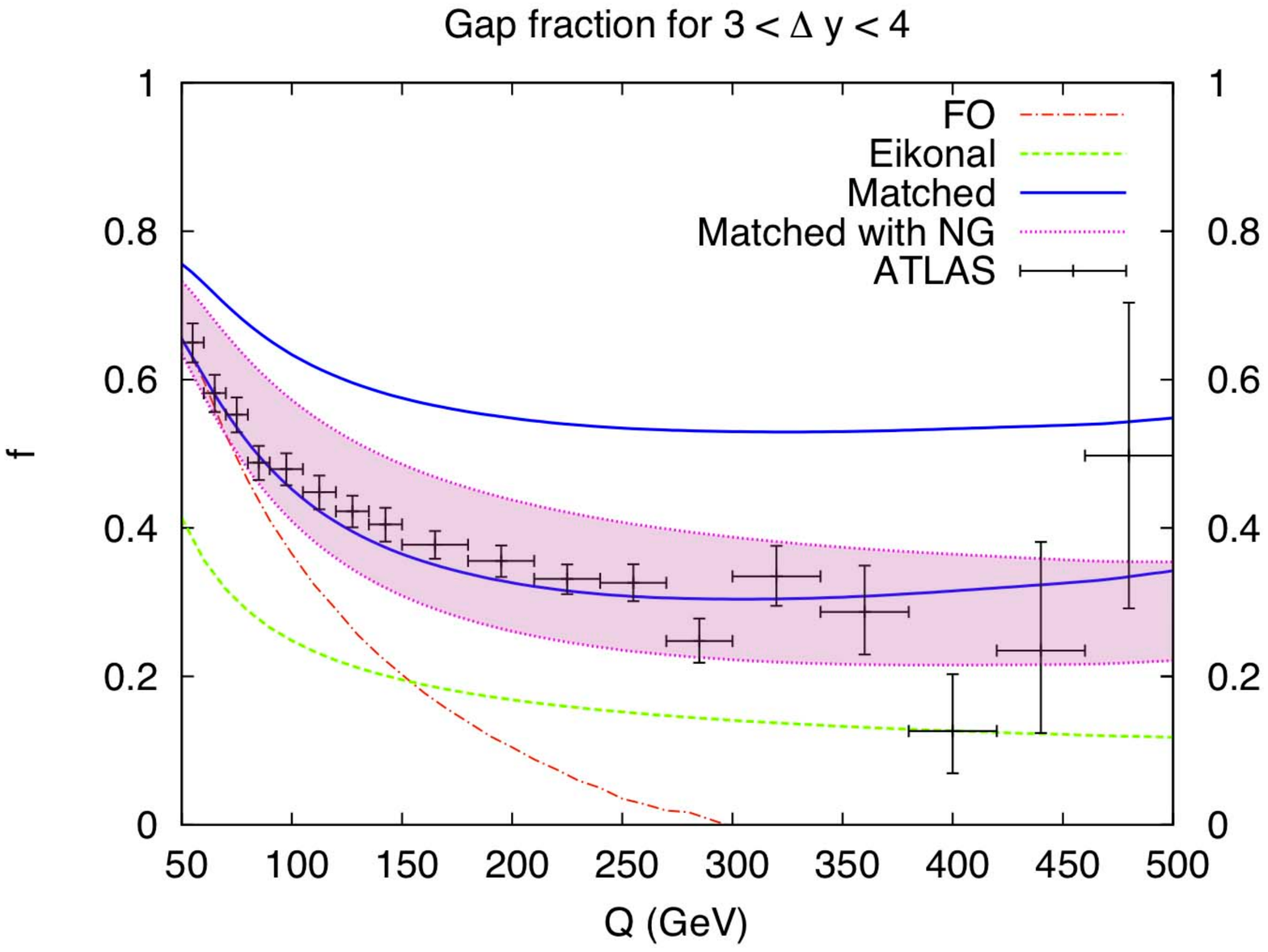}
\includegraphics[width=0.49\textwidth, clip]{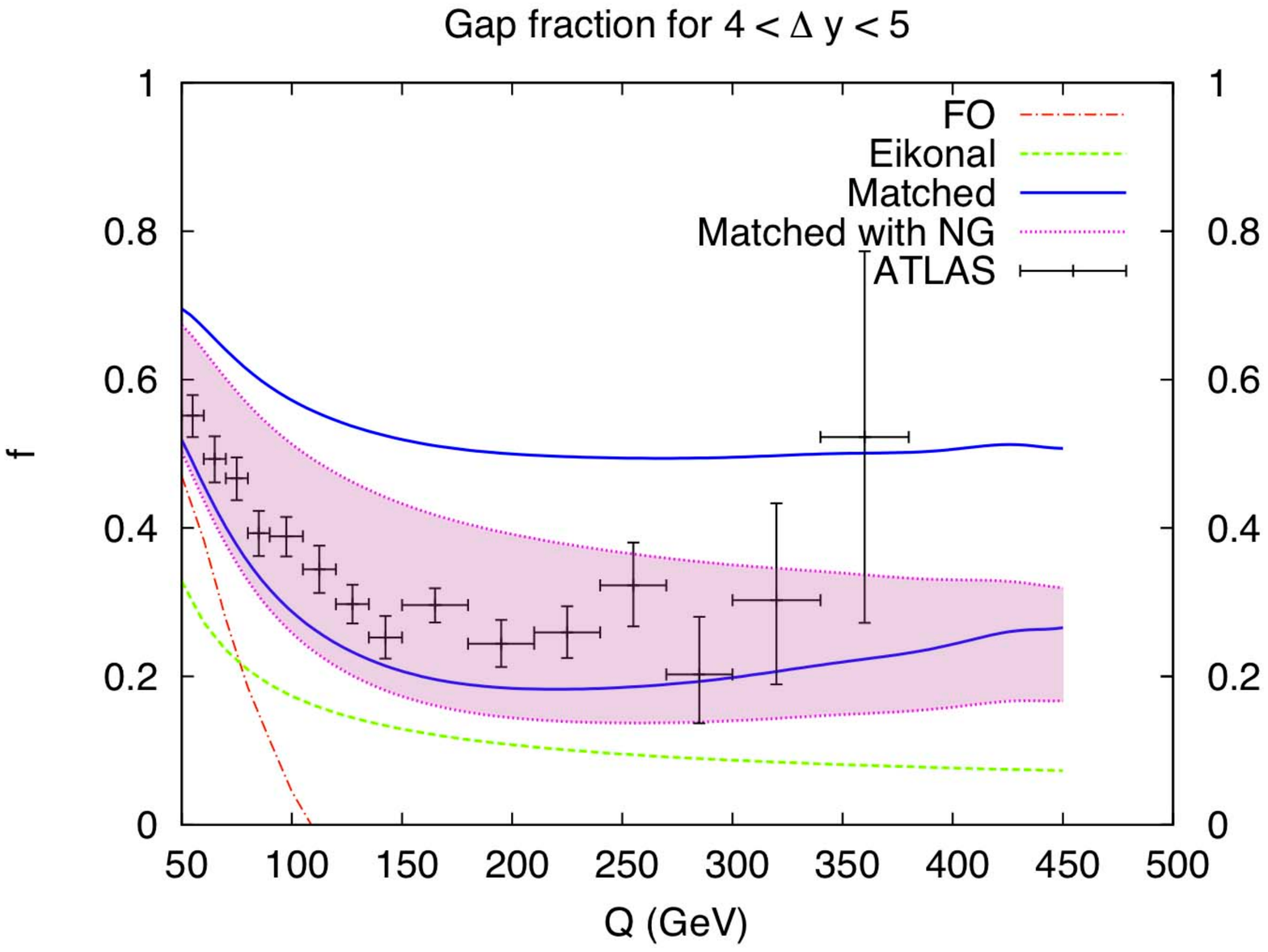}
\caption{The matched gap fraction as a function of the transverse momentum $Q$ in different rapidity bins. }\label{fig:match2}
\end{center}
\end{figure}

The plots in Figs.~\ref{fig:match1} and \ref{fig:match2} show the gap fraction as a function of $Q$ in four different rapidity bins.
The black crosses are the data points measured by the ATLAS collaboration~\cite{atlasveto} with the gap defined by the two highest $p_T$ jets.
 The LO calculation (dash-dotted red line) is clearly only sensible in the first rapidity bin and for $\Delta y > 2$, it decreases very rapidly as a function of $Q$ and eventually becomes negative. This unphysical behaviour is driven by a large logarithmic term $\sim \as \Delta y \ln \frac{Q}{Q_0}$, which needs to be resummed. The eikonal resummation (dashed-green line) restores the physical behaviour but, as we have previously discussed, completely ignores the issue of energy-momentum conservation and produces too small a gap fraction. Our matched curves (blue band), with the inclusion of non-global logarithms (magenta band), do seem to capture most of the salient physics. However, our results are affected by large theoretical uncertainties due to the fact that the calculation is accurate only at the leading logarithmic level.

\section{Conclusions and Outlook}
We have discussed jet vetoes as a probe of colour flow in hard scatterings.  Perturbative choices for the veto scale $Q_0$ reduce the influence of the underlying event  and yield good agreement between the dijet data and resummed perturbation theory, although theoretical predictions are affected by large uncertainties.  Nevertheless, the situation can be improved by matching to NLO~\cite{nlojet}.

Having validated the theoretical framework with a process with a relatively simple final state, our interest now lies in exploring more complicated processes, within and beyond the Standard Model.
The ATLAS collaboration have recently measured the $t \bar{t}$ cross-section in the presence of a jet veto~\cite{ttbexp}. The process $Z+ 2\,{\rm jets}\,+\,{\rm veto}$ is also interesting~\cite{Zjj} and we plan to study this in the near future as an important step towards understanding Higgs production in association with two jets~\cite{Hjj}.
 
 \begin{footnotesize}

  \end{footnotesize}
\end{document}